\documentstyle[12pt]{article}
\newcommand{\bce}{\begin{center}}
\newcommand{\ece}{\end{center}}
\newcommand{\bea}{\begin{eqnarray}}
\newcommand{\eea}{\end{eqnarray}}
\newcommand{\be}{\begin{equation}}
\newcommand{\ee}{\end{equation}}
\newcommand{\bd}{\begin{displaymath}}
\newcommand{\ed}{\end{displaymath}}
\newcommand{\bit}{\begin{itemize}}
\newcommand{\eit}{\end{itemize}}
\newcommand {\ben}{\begin{enumerate}}
\newcommand{\bfx}[1]{\mbox{\boldmath $#1$}}
\newcommand{\een}{\end{enumerate}}

\begin{document}
\bce
{
\bf
Rates for the reactions
$\overline{p}p\rightarrow\pi\phi$ and $\gamma\phi$
}

\vspace{0.5cm}

M.P. Locher$^a$,
Y. Lu$^a$ and B.S. Zou$^b$ \\
\vspace{0.3cm}
$^a$ Paul Scherrer Institute, CH--5232 Villigen PSI, Switzerland\\
$^b$ Queen Mary and Westfield College, London E1 4NS, UK
\ece

\vspace{1cm}
\begin{abstract}

We study $\overline{p}p$ annihilation at rest
into $\pi\phi$ and $\gamma\phi$. Rescattering
by $\overline{K^*}K+K^*\overline{K}$ and $\rho^{+}\rho^{-}$ for
$\overline{p}p\rightarrow\pi\phi$ states is sizable,
of order $(0.90\, {\rm to}\,2.6)\times 10^{-4}$
in the  branching ratio, but smaller than experiment.
For $\overline{p}p\rightarrow\gamma\phi$ the
rescattering contributions are negligible, but
the $\gamma\phi$ channel is well explained by a $\rho\phi$ intermediate state
combined with vector meson dominance.

\end{abstract}
\newpage



\indent

{\noindent \bf 1. Introduction}

In the present paper we examine the rates for
$\overline{p}p\rightarrow\pi\phi$ and $\overline{p}p\rightarrow\gamma\phi$
at rest by calculating the contributions from
two-meson-intermediate states. The calculation
of rescattering contributions for
$\overline{p}p\rightarrow\phi\phi$ in flight from Ref.~\cite{LZL} is
extended to the present reactions. In addition, for
$\overline{p}p\rightarrow\gamma\phi$, vector
meson dominance mechanisms are evaluated.

\vspace{0.5 cm}
{\noindent \bf 2. The reaction $\overline{p}p\rightarrow\pi\phi$}

We start by evaluating the two meson rescattering contributions.
Important contributions are expected from intermediate states containing a
$K^{\ast}$ and a $K$, as shown in Fig.~1a
which describes the
contribution to $\overline{p}p\rightarrow
\pi\phi$. Similarly the contributions from $\rho^+\rho^-$ intermediate
states (Fig.~1b) are expected to be large since $\overline{p}p\rightarrow\rho^+
\rho^-$ is large, compensating the smallness of the $\phi\rightarrow\pi\rho$
vertex.
Following the formalism of Ref.~\cite{LZL}, we describe for definiteness
the amplitude
for $\overline{p}p(^3 S_1)\rightarrow
K^{\ast}\overline{K}\rightarrow\pi\phi$
\be
T=\int\;
\frac{d^4 p_{1}}{(2\pi)^4}
T_{S}\,
T_{\phi K\overline{K}}\, T_{K^{\ast}K\pi}\, G_{1} G_{2} G_{3}\,.
\ee
Here the amplitudes $T_{\phi K\overline{K}}$, $T_{K^{\ast}K\pi}$ and $T_{S}$
are for
the $\phi K \overline{K}$, $K^{\ast}K\pi$
and $\overline{p}p(^3 S_1)
\rightarrow K^{\ast}\overline{K}$ three body vertices, respectively.
The meson propagators are denoted by $G_i$ ($i=1,2,3$).
In the following we evaluate the part of the amplitude $T$ of Eq.~(1)
where  $K^{*}$  and $\overline{K}$ propagate on the mass shell.
All the amplitudes in Eq.~(1)
are then directly related to experimental
information\cite{PDB,Conforto,Bettini}. Similar equations hold for
$\rho^+\rho^-$ intermediate states.

We define
the vertices in Eq.~(1) in relativistic notation as follows:
\begin{itemize}
\item The amplitude for the annihilation $\overline{p}p(^3 S _1)
\rightarrow K^{\ast}\overline{K}$ with the smallest number of derivatives
is
\be
T_{S}=g_{S}\, \epsilon^{\alpha\beta\gamma\delta}
\,e_{\alpha}(\overline{p}p,\; m_1)
\,p_\beta(\overline{p}p)
\,e_{\gamma}(K^{\ast},\; m_2)
\,p_{\delta}(K^\ast)\;,
\ee
where $\epsilon^{\alpha\beta\gamma\delta}$ is the rank--4
antisymmetric tensor,
$e_{\alpha}(\overline{p}p,\; m_1)$ and
 $e_{\beta}(K^{\ast},\; m_2)$ the polarization vectors
of the initial state and $K^\ast$ meson, respectively.
The parameter $g_S$
is  related to the $\overline{p}p(^3 S_1)\rightarrow K^\ast\overline{K}$
transition strength by\footnote{We denote the magnitudes
of three--momenta by capital letters.}
\be
\Gamma_{\overline{p}p(^3 S _1)\rightarrow K^{\ast}\overline{K}}=\frac{1}
{12\pi}\,g_{S}^2\, P_{K^\ast\overline{K}}^3\,.
\ee
\item The amplitudes for the $K^{*}$ and $\phi$ decays are
parametrized as follows,
\be
T_{\phi K\overline{K}}
=g_{\phi K\overline{K}}\,e(\phi,\;m_3)\cdot(2p_2-q_2),
\ee
\be
T_{K^{\ast} K\pi}=g_{K^{\ast} K\pi}\,e(K^{\ast},\;m_2)
\cdot(2q_1-p_1).
\ee
The coupling constants are related to the vector meson decay
widths by
\be
\Gamma_{\phi K\overline{K}}=\frac{1}{6\pi}\frac{g_{\phi K\overline{K}
}^2\,P^3_{\overline{K}K(\phi)}}
{M_{\phi}^2}
\ee
and
\be
\Gamma_{K^{\ast}
K\pi}=\frac{1}{6\pi}\frac{g_{K^{\ast} K\pi}^2\, P^3_{K\pi(K^{\ast})}}
{M_{K^{\ast}}^2}
\ee
where $P_{\overline{K}K(\phi)}$ and $P_{K\pi(K^\ast)}$
are the magnitudes of the 3-momenta of the decay products
in the CM frames of the vector mesons.
\end{itemize}

Next we calculate the amplitude Eq.~(1)
for the on-shell $K^{\ast +}K^{-}$ intermediate state:
\be
T^{\rm on-shell}_{\overline{p}p(^3 S _1)\rightarrow K^{\ast +}K^{-}\rightarrow
\phi\pi}
=
\frac{g_S \,g_{\phi K\overline{K}}\, g_{K^{\ast} K\pi}\, P}{32\, \pi^2}
\int d\Omega_{\bf p}
\frac
{
{\bfx{e}}(\overline{p}p,\;m_{1})\cdot(
{\bf p}\times {\bf q}) \; e(\phi,\; m_3)\cdot (2p_2-q_2)}
{({\bf p}-{\bf q})^2+m_{K}^2-(E_{\phi}-E_{K^{-}})^2}
\ee
where the momenta in Fig.~1a are denoted by
\bea \nonumber
&&p_{1}=(E_{K^{\ast +}},-{\bf p}),\; \;\; p_{2}=(E_{K^{-}},{\bf p})\\ \nonumber
&&q_{1}=(E_{\pi},-{\bf q}),\;\; \;\;\;q_{2}=(E_{\phi}, {\bf q}) \\
&&p_{3}=p_1-q_1=q_2-p_2\;.
\eea
The resulting decay width of $\overline{p}p(^3 S_1)\rightarrow\pi\phi$
is
\be
\Gamma_{\overline{p}p(^3 S _1)\rightarrow K^{\ast +}K^{-}\rightarrow
\pi\phi}
=\frac{1}{8\pi}\frac{Q}{M_{\overline{p}p}^2}\frac{1}{3}
\sum_{m_1 m_3} |
T^{\rm on-shell}_{\overline{p}p(^3 S _1)\rightarrow K^{\ast +}K^{-}\rightarrow
\pi\phi}
|^2\;.
\ee
After summing over the spins and doing the angle integrations, we arrive at
\be
\sum_{m_1 m_3} |
T^{\rm on-shell}_{\overline{p}p(^3 S _1)\rightarrow K^{\ast +}K^{-}\rightarrow
\pi\phi}
|^2
=
\frac{g_S^2 \,g^2_{\phi K\overline{K}}
\, g^2_{K^{\ast} K\pi}\, P^4}{512 \,\pi^2}C_0^2\;.
\ee
Here the dimensionless function $C_0$ obtained from the
loop integration is
\be
C_0=2z+(1-z^2)\ln\left| \frac{z+1}{z-1}\right|
\ee
with $z=(P^2+Q^2 +m_{K}^2-(E_{\phi}-E_{K^{-}})^2)/(2P Q)$.

Collecting all factors, we obtain
\be
\frac
{
\Gamma_{\overline{p}p(^3 S _1)\rightarrow K^{\ast +}K^{-}\rightarrow
\phi\pi}
}
{
\Gamma_{\overline{p}p(^3 S _1)\rightarrow K^{\ast +}K^{-}}
}
=
\frac{9}{256}
\frac
{
P\,Q\,M_{\phi}^2 \,M_{K^{\ast}}^2 \;\Gamma_{\phi\rightarrow K^{+}K^{-}}
\;\Gamma_{K^{\ast+}\rightarrow K^{+}\pi_0}
}
{
M_{\overline{p}p}^2\, P^3_{K^{+}K^{-}(\phi)}
\,P^3_{K^+\pi^{0}(K^{\ast +})}
}\,C_0^2\,.
\ee
In Eq.~(13) we have only included the  contribution
from intermediate $K^{\ast +}K^{-}$. Note that $\Gamma_{K^{*+}
\rightarrow K^+\pi^0}$ is $1/3$ of the total $K^{*+} $
width and $\Gamma_{\phi\rightarrow K^{+}K^{-}}	$ is $1/2$ of
the total $\phi$ width. Since our reaction is
restricted to isospin one and negative charge conjugation,
the contributions from $K^{\ast +}K^{-}$, $K^{\ast -}K^{+}$,
$K^{\ast 0}\overline{K}^0$  and
$\overline{K}^{\ast 0}K^{0}$ are all equal on the amplitude level.
Therefore the total width of
$\overline{p}p(^3 S _1)\rightarrow\pi\phi$ from
all $ K^{\ast}K$ states is
\be
\Gamma_{\overline{p}p(^3 S _1)\rightarrow
K^{\ast}\overline{K}+\overline{K^\ast} K
\rightarrow\pi\phi}=16\;
\Gamma_{\overline{p}p(^3 S _1)\rightarrow K^{-}K^{\ast +}\rightarrow\pi\phi}
\ee
and
\be
\Gamma_{
\overline{p}p(^3 S _1)\rightarrow K^{\ast}\overline{K}+\overline{K^\ast} K}
=4\;
\Gamma_{\overline{p}p(^3 S _1)\rightarrow K^{\ast+}K^{-}}\,.
\ee
The total branching ratio in the on-shell approximation is then
\be
\frac
{
\Gamma_{\overline{p}p(^3 S _1)\rightarrow K^{\ast}\overline{K}+
\overline{K^\ast}K
\rightarrow
\pi\phi}
}
{
\Gamma_{\overline{p}p(^3 S _1)\rightarrow K^{\ast}\overline{K}+
\overline{K^\ast} K
}
}
=
4\;\frac
{
\Gamma_{\overline{p}p(^3 S _1)\rightarrow K^{-}K^{\ast +}\rightarrow
\phi\pi}
}
{
\Gamma_{\overline{p}p(^3 S _1)\rightarrow K^{-}K^{\ast +}}
}\,.
\ee
The experimental branching ratio is $BR[
\overline{p}p(^3 S _1,\,I=1,C=-1)
\rightarrow K^{\ast}\overline{K}+\overline{K}^\ast K)]
\geq (23.4\pm 2.0)\times10^{-4}$ from Ref.~\cite{Conforto}.
This number is a lower limit since the projection
on $I=1$, $I_3=0$, $C=-1$ states is experimentally incomplete\cite{Conforto}.
The missing contributions are of order $30\%$. Taking the lower limit we
calculate the
branching ratio of $\pi\phi$ from Eq.~(16). The result
is $BR(\pi\phi)=\Gamma_{\pi\phi}/\Gamma_{\rm tot}=0.6\times 10^{-4}$.
Among similar lines we have also estimated the contributions from
$\rho^+\rho^-$ and $\pi^+\pi^-$\cite{AmMy}  intermediate states.
The branching ratio for $\overline{p}p\rightarrow\rho^{+}\rho^{-}$
is experimentally not known. Conservative interpolations indicate
a range of $10$ to $40\times 10^{-3}$. If we
use the theoretical estimate $23.6\times 10^{-3}$ from \cite{CLZ}
 the $\pi\phi$ branching ratio is $0.8\times 10^{-4}$.
The $\rho^{+}\rho^{-}$ contribution is thus expected to be bigger than
$K^{\ast}K$. The corresponding estimate for the $\pi\phi$
branching ratio from $\pi^{+}\pi^{-}$ states is $1.0\times 10^{-6}$
which is negligible.
So far our results are for intermediate states propagating
on-shell.
However, the kaon with momentum $p_3$ in Fig.~1a is off mass-shell and
so is the exchanged pion in Fig.~1b for the $\rho^{+}\rho^{-}$ diagram. We
introduce a monopole form factor $F(p^2_3)=(\Lambda^2-m_{K}^2)/
(\Lambda^2-p_3^2)$ at the
  $\phi KK$ and $K^{\ast}K\pi$ vertices in Fig.~1a.
For the $\rho^{+}\rho{-}$ case
just exchange $m_{K}$ with $m_{\pi}$.
Varying $\Lambda$ from $1.2$ GeV to $2.0$ GeV, the corresponding
variation of the $\pi\phi$
branching
ratio from $K^\ast K$ states is  ($0.14$ to $0.33)\times
10^{-4}$.  Similarly the corresponding range for the $\rho^{+}\rho^{-}$
exchange is $(0.31$ to $0.52)\times10^{-4}$. The reduction is smaller in
this case since the exchanged pion is almost on shell. In addition we
 expect a contribution of similar size\cite{MZ}
from $K^{\ast}K$ or $\rho^{+}\rho^-$ states propagating
off-shell.
Adding this contribution, considering the range induced by the variation
of the form factors  and the experimental uncertainties we obtain the
following range for the
total rescattering contribution
\be
BR(\pi\phi)=(0.90\,{\rm to}\,2.6)\times 10^{-4}
\ee
while the
experiment\cite{Braune} gives $(4.0\pm0.8)\times 10^{-4}$.
Given the theoretical uncertainties the rescattering contributions
are thus sizable but somewhat small when compared to experiment.

\vspace{0.5 cm}
{\noindent \bf 3. The reaction $\overline{p}p\rightarrow
\gamma\phi$}

\vspace{0.3 cm}
{\noindent \bf a) Rescattering contribution}

We calculate
$K^{\ast}K$
on--shell rescattering for $\overline{p}p(^1 S_0)\rightarrow
\gamma\phi$, corresponding to Fig.~1c.
The procedure is very similar to the
$\pi\phi$ case, with the result
\be
\frac
{
\Gamma_{\overline{p}p(^1 S _0)\rightarrow K^{\ast}\overline{K}
+\overline{K^\ast }K  \rightarrow
\gamma\phi}
}
{
\Gamma_{\overline{p}p(^1 S _0)\rightarrow
K^{\ast}\overline{K}
+\overline{K^\ast} K }
}
=
\frac{9}{64}
\frac
{
P\,Q\,M_{\phi}^2 \,M_{K^{\ast}}^2 \;\Gamma_{\phi\rightarrow K^{+}K^{-}}
\;\Gamma_{K^{\ast+}\rightarrow K^{+}\gamma}
}
{
M_{\overline{p}p}^2\, P^3_{K^{+}K^{-}(\phi)}
\,P^3_{K^+\gamma(K^{\ast +})}
}\,C_0^2\,.
\ee
Here the function $C_{0}$ is the same as in Eq.~(12) with
suitable replacements of momenta and masses.
 Using the value $12.0\times10^{-4}$ for the
branching ratio of
$\overline{p}p(^1 S_0)\rightarrow K^{\ast}\overline{K}+\overline{K^\ast} K$
from Ref.~\cite{Conforto},
we have
$BR[\overline{p}p(^1 S_0)\rightarrow \gamma\phi]=1.7\times 10^{-7}$.
Including a form factor $F(p^2_3)=(\Lambda^2-m_{K}^2)/
(\Lambda^2-p_3^2)$ at the $\phi KK$ and $K^{\ast}K\gamma$ vertices and
varying $\Lambda$ from $1.2$ GeV to $2.0$, the corresponding
variation of the $\gamma\phi$ branching ratio
is $3.0\times 10^{-8}$
to $9.4\times 10^{-8}$. Comparing with the experimental value
of
$BR[\overline{p}p(^1 S_0)\rightarrow \gamma\phi]=(1.2\pm0.3)\times 10^{-5}$, we
find the on--shell contribution from $K^{\ast}K$ is about two to
three orders of magnitude smaller.

\vspace{0.3 cm}
{\noindent \bf b) Contribution from $\rho\phi$ via vector meson dominance}

Vector meson dominance has been
used to successfully estimate the branching ratios of $\overline{p}p$
to $\gamma\gamma$, $\gamma\pi$,$\gamma\eta$, $\gamma\omega$ and
$\gamma\eta'$\cite{Amsler1}. For our reaction vector meson dominance
applied to a $\rho\phi$ intermediate state is illustrated in Fig.~2a.
In order to evaluate this diagram, we parametrize the amplitude of
$\overline{p}p(^1 S_0)\rightarrow\rho\phi$ by
\be
M_A=g\,\epsilon^{\alpha\beta\mu\nu}e_{\alpha}(\rho)
p_\beta (\rho)e_{\mu}(\phi)p_\nu (\phi)
\ee
where
$g$ is the strength of the annihilation. The constant $g$
is related to the partial width of
$\overline{p}p(^1 S_0)\rightarrow\rho\phi$ by
\be
\Gamma_{\overline{p}p(^1 S_0)\rightarrow\rho\phi}
=\frac{g^2}{4\pi}P_1^3\,.
\ee
Here $P_1$ is the 3-momentum of $\rho\phi$ in the $\overline{p}p$
CM frame.

The invariant amplitude of $\overline{p}p(^1 S_0)\rightarrow
\rho\phi\rightarrow\gamma\phi$ can then be written as
\be
T=g\, \gamma_{\rho\gamma}\, \epsilon^{\alpha\beta\mu\nu}
\,e_\alpha(\phi)\,
p_\beta (\phi)\, G_{\mu\delta}(\rho)\,
e^{\delta}(\gamma)\,p_\nu(\gamma)
\ee
where $G_{\mu\delta}(\rho)$ is the propagator of the
$\rho$ meson with a 4-momentum corresponding to that of the photon.
The effective $\gamma\rho$
coupling constant $\gamma_{\rho\gamma}$ can
be expressed by the universal constant $f_\rho$ as
$\gamma_{\rho\gamma}=eM^2_{\rho}/f_\rho$ \cite{Sakurai}.
The partial width  is therefore
\be
\Gamma_{\overline{p}p(^1 S_0)\rightarrow
\rho\phi\rightarrow\gamma\phi}
=\frac{g^2}{4\pi}\frac{e^2}{f_{\rho}^2}Q_1^3
\ee
with $Q_1$ being the three momentum of $\gamma\phi$ in the
$\overline{p}p$ CM frame.

Combining Eq.~(20) and (22), we have
\be
\frac
{
\Gamma_{\overline{p}p(^1 S_{0} )\rightarrow
\rho\phi\rightarrow\gamma\phi}
}
{
\Gamma_{\overline{p}p(^1 S_{0} )\rightarrow
\rho\phi}
}
=\frac{e^2}{f_{\rho}^2}\frac{Q_1^3}{P_1^3}
\,.
\ee
With $e^2/4\pi=1/137$, $f^2_\rho/4\pi=2.5$ from \cite{Sakurai} and
$BR[\overline{p}p(^1 S_0)\rightarrow
\rho\phi]=(3.4\pm1.0)\times10^{-4}$ from Ref.~\cite{AmMy}, the contribution
from $\rho$ dominance is
\be
BR[\overline{p}p(^1 S_0)\rightarrow
\rho\phi\rightarrow\gamma\phi]=1.27\times10^{-5}\,.
\ee
This is very close to the preliminary result  of
the order of $1.0\times 10^{-5}$ from the Crystal Barrel Collaboration
\cite{Fassler,Amsler2}.

Because the $\omega\gamma$ coupling constant is one third
of the $\rho\gamma$ value(see \cite{Meissner}, e.g.), the $\omega$
contribution to $\gamma\phi$ is smaller. With
$BR[\overline{p}p(^1 S_0)\rightarrow
\omega\phi]=(5.3\pm2.2)\times 10^{-4}
$ from Ref.~\cite{AmMy}, we obtain
\be
BR[\overline{p}p(^1 S_0)\rightarrow
\omega\phi\rightarrow\gamma\phi]=2.8\times10^{-6}\,.
\ee
The interference
between the $\rho$ and $\omega$ amplitudes
is destructive from experimental analysis\cite{Amsler1}. The reduction
is however not significant in our case.

We have also applied vector meson dominance to $\overline{p}p(^1 S_0)
\rightarrow\gamma\omega$. The experimental
rates for the intermediate states are
$BR[\overline{p}p\rightarrow\omega\rho]=1.91\times10^{-2}$
 \cite{AmMy,Dover} and  $BR[\overline{p}p\rightarrow\omega\omega
=3.32\times10^{-2}$
\cite{Amsler4}. Using Eq.~(23) with suitable replacements, we obtain
$BR[\overline{p}p\rightarrow\omega\rho\rightarrow\omega\gamma]=1.76\times
10^{-4}$ and
$BR[\overline{p}p\rightarrow\omega\omega\rightarrow\omega\gamma]=0.72\times
10^{-4}$. In the case of destructive interference the resulting
branching ratio is
 $BR[\overline{p}p\rightarrow\gamma
\omega]=2.3\times 10^{-5}$
while the experiment\cite{Amsler1} gives $(6.8\pm1.8)\times10^{-5}$,
indicating preference for destructive interference.

\vspace{0.5cm}
{\noindent \bf 4. Concluding remarks}

We have evaluated the leading rescattering contributions for the reaction
$\overline{p}p\rightarrow\pi\phi$. We found that the contributions from
$\rho^{+}\rho^-$ and $K^{\ast}K$ intermediate states are largest and similar in
size with a combined contribution of order $(0.9\,{\rm to}
\,2.6)\times10^{-4}$ for the
branching  ratio\footnote{After completing this calculation we have received
the preprint \cite{buzatu} where rescattering contributions for
$\overline{p}p\rightarrow\pi\phi$ and similar reactions have been
evaluated. The result obtained for the on-shell
$K^\ast K$ contribution to $\pi\phi$, e.g., is too big.
A large enhancement has been traced to an error in \cite{buzatu} for the
loop integration, compare with our Eq.~(12).}
of this OZI suppressed reaction. This
contribution is sizable but it does not quite reach experiment.
For the reaction $\overline{p}p\rightarrow\gamma\phi$ the contribution
from $K^{\ast}K$ rescattering is negligible (of
order $ 10^{-7}$ in the branching ratio). However, the $\rho\phi$
and $\omega\phi$ intermediate states combined with vector meson
dominance, see Eq.~(24) and (25), easily explain the experimental
rate.

\vspace{0.5cm}

{\noindent \bf Acknowledgement}
\vspace{0.3cm}

We are grateful to Claude Amsler  for numerous discussions
and information on experimental results. We thank
M.G.~Sapozhnikov for drawing our attention to Ref.~\cite{buzatu}.
We are indebted to Roland Rosenfelder for
a critical reading of the manuscript and
David Bugg for discussions.

\newpage
\noindent
{\large \bf Figure captions}
\vskip 2 true cm

{\bf Fig.~1: }a. Triangle diagram for $\overline{p}p\rightarrow
{K^\ast}K\rightarrow
\pi\phi$; b. Triangle diagram for $\overline{p}p\rightarrow
\rho^+\rho^-\rightarrow
\pi\phi$; c. Triangle diagram for $\overline{p}p\rightarrow
{K^\ast}K\rightarrow
\gamma\phi$.

{\bf Fig.~2: }a. $\rho$ dominance contribution to
$\overline{p}p\rightarrow
\gamma\phi$; b. $\omega$ dominance contribution to
$\overline{p}p\rightarrow
\gamma\phi$.

\end{document}